%% file: conference_101719.tex
\documentclass[conference]{IEEEtran}
\IEEEoverridecommandlockouts
\usepackage{cite}
\usepackage{amsmath,amssymb,amsfonts}
\usepackage{algorithmic}
\usepackage{graphicx}
\usepackage{textcomp}
\usepackage{xcolor}
\usepackage{booktabs}
\usepackage{subcaption}
\usepackage{longtable}
\usepackage{multirow}
\usepackage{hyperref}

\def\BibTeX{{\rm B\kern-.05em{\sc i\kern-.025em b}\kern-.08em
    T\kern-.1667em\lower.7ex\hbox{E}\kern-.125emX}}
    
\begin{document}

\title{JamShield: A Machine Learning Detection System for Over-the-Air Jamming Attacks}

\author{
\IEEEauthorblockN{
Ioannis Panitsas\textsuperscript{*‡}, 
Yagmur Yigit\textsuperscript{†‡}, 
Leandros Tassiulas\textsuperscript{*}, 
Leandros Maglaras\textsuperscript{†}, 
Berk Canberk\textsuperscript{†}}

\IEEEauthorblockA{
\textsuperscript{*}Department of Electrical and Computer Engineering, Yale University\\}
\IEEEauthorblockA{
\textsuperscript{†}School of Computing, Engineering and The Built Environment, Edinburgh Napier University\\}
}

\maketitle

\begingroup\renewcommand\thefootnote{\textsuperscript{‡}}
\footnotetext{These authors contributed equally to this work.}
\endgroup

\begin{abstract}
Wireless networks are vulnerable to jamming attacks due to the shared communication medium, which can severely degrade performance and disrupt services. Despite extensive research, current jamming detection methods often rely on simulated data or proprietary over-the-air datasets with limited cross-layer features, failing to accurately represent the real state of a network and thus limiting their effectiveness in real-world scenarios. To address these challenges, we introduce \textit{JamShield}, a dynamic jamming detection system trained on our own collected over-the-air and publicly available dataset. It utilizes hybrid feature selection to prioritize relevant features for accurate and efficient detection. Additionally, it includes an auto-classification module that dynamically adjusts the classification algorithm in real-time based on current network conditions. Our experimental results demonstrate significant improvements in detection rate, precision, and recall, along with reduced false alarms and misdetections compared to state-of-the-art detection algorithms, making \textit{JamShield} a robust and reliable solution for detecting jamming attacks in real-world wireless networks.
\end{abstract}


\section{Introduction}

Wireless networks have become a vital component of modern communication systems due to the numerous services they provide to users. Security has always been an important aspect of network design, especially for wireless networks, which have a larger attack surface due to the shared access medium and the limited spectrum \cite{b1}. Many wireless technologies are vulnerable to physical-layer security threats, particularly \textit{jamming attacks}, which compromise network availability. Jamming attacks are a form of denial-of-service attack that interfere with wireless radio communications by overlaying legitimate signals with a noise signal of significantly higher power, thereby degrading the Signal-to-Noise Ratio (SNR) through the use of a jamming signal source \cite{b1}. These attacks can have devastating results in critical systems, potentially disrupting communication networks used in emergency response, public safety, and other essential
services. With the evolution of Software-Defined Radios (SDRs), these attacks can now be executed using low-cost devices with minimal complexity, thanks to the advanced signal processing features they offer. For instance, a \$20 USB dongle can be programmed to transmit malicious interference signals, covering a 20 MHz bandwidth below 6 GHz and transmitting up to 100 mW of power \cite{b2}. More advanced off-the-shelf SDR devices, such as the Universal Software Radio Peripheral (USRP), provide even greater power and flexibility for launching jamming attacks.

Given this ease of executing jamming attacks, there is an urgent need for robust detection mechanisms to mitigate malicious interference. Over the past decade, Machine Learning (ML) has been effectively applied to complex challenges, including network security \cite{b3}, by identifying patterns and features linked to malicious behavior.  Building on this foundation, several research
studies have focused on developing learning-based techniques for the detection and classification of jamming attacks [4-12]. More specifically, various ML frameworks were implemented in \cite{b4}, \cite{b7}, \cite{b12} to identify and classify jamming attacks in both 802.11 and vehicular networks, evaluated using simulated data. The primary performance metrics employed for jamming detection included Packet Delivery Ratio (PDR), Received Signal Strength (RSS), and SNR.  In addition, the authors in \cite{b6} and \cite{b8} implemented a similar framework in IoT wireless networks, utilizing over-the-air RSS values for analysis and evaluation. Finally, the authors in \cite{b5}   proposed a ML scheme for detecting and classifying jamming attacks in both indoor and outdoor environments within 802.11 networks, leveraging features from both the physical and application layer.

Despite these valuable contributions, several research gaps still exist in the literature. First, many approaches primarily evaluate their models using simulated data \cite{b4, b7, b9, b12}, which may not accurately reflect the real network state, thus limiting the generalizability of their findings. Second, while some studies incorporate over-the-air datasets \cite{b5, b6, b8, b10, b11}, they often rely on a limited set of features predominantly focusing on metrics like RSS and SNR, which cannot comprehensively capture network performance under jamming conditions. Finally, most available datasets are proprietary and inaccessible  \cite{b6}, \cite{b5}, \cite{b10}, \cite{b11}. Although some open-source datasets exist \cite{b8}, \cite{b13}, they typically provide limited features, samples, and evaluations across a few types of jammers.

To address these gaps, our work introduces a robust dynamic jamming detection system, called \textit{JamShield}. The system employs a hybrid feature selection approach to minimize the number of network features for accurate jamming detection, while also incorporating an auto-classification module that dynamically adapts to real-time network conditions. Evaluation results showcase \textit{JamShield's} superior performance in detection rate, precision, and recall, along with a reduction in false alarms and misdetections compared to state-of-the-art jamming detection algorithms. Overall, the main contributions of this work are:
\begin{itemize}
   \item We collect an extensive set of cross-layer 802.11 features from our testbed, capturing diverse network configurations, traffic patterns, and jamming scenarios to ensure broad applicability.
   \item We release a publicly available open-source dataset \cite{b16}, \cite{dataport} to foster further research in this area.
   \item We propose a hybrid feature selection method that combines Principal Component Analysis (PCA) and Mutual Information (MI) through a weighted voting mechanism, ensuring the most impactful features are selected for real-time jamming detection while reducing dimensionality and maintaining high accuracy.
   \item We design \textit{JamShield}, a network-aware jamming detection system with an Automated Classification Module (AutoCM) that dynamically selects the best classification algorithm in real-time, optimizing detection accuracy while minimizing overhead.
\end{itemize}

\section{Experimental Setup}

\begin{figure*}[hbt!]
    \centering
    \begin{subfigure}[b]{0.3\textwidth}
        \centering
        \includegraphics[width=\textwidth, height=4.5cm]{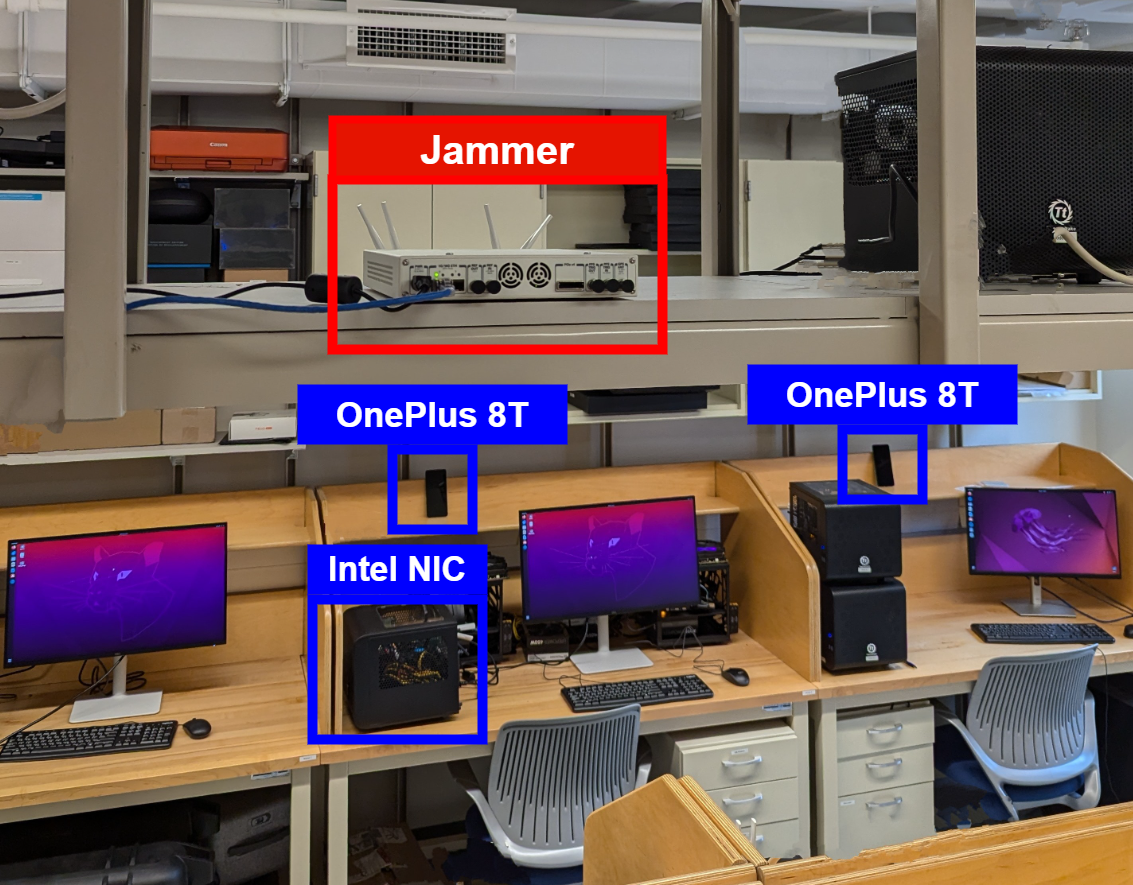}
        \caption{Position of the jammer (LOS scenario)}
        \label{fig:position_jammer}
    \end{subfigure}
     \hspace{0.05\textwidth}
     \begin{subfigure}[b]{0.3\textwidth}
        \centering
        \includegraphics[width=\textwidth, height=4.5cm]{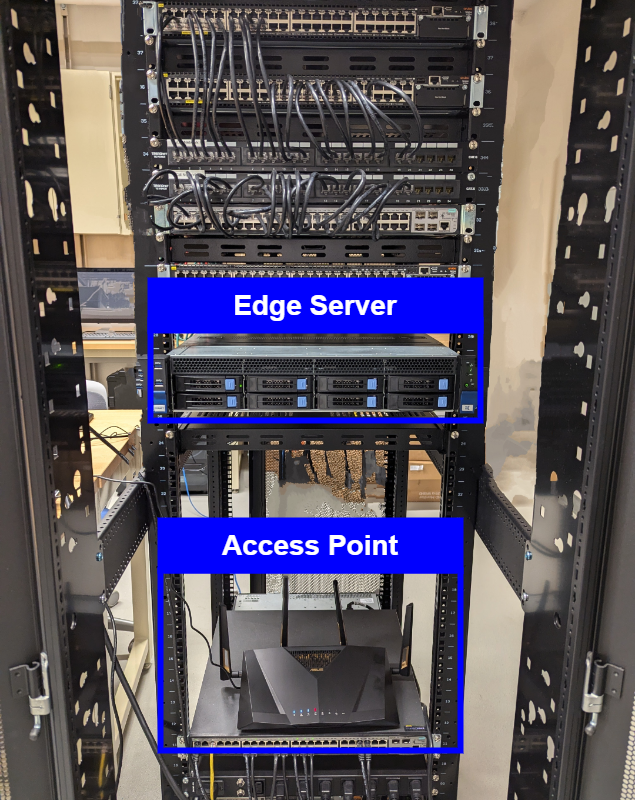}
        \caption{Wireless AP and edge server}
       \label{fig:edge_server}
    \end{subfigure}
    \caption{Overview of the experimental setup, including the jammer position, wireless AP, and edge server.}
    \label{fig:experimental_set_up}
\end{figure*}

\subsection{Hardware Setup}
Our experimental setup is deployed in an area of 80 m² and includes three fixed-position wireless nodes, two OnePlus 8T smartphones, one wireless access point (AP), one USRP X310, and one edge server, as illustrated in Fig. 1. The placement of each device is shown in Fig. 2 for two different scenarios, which are further detailed in Section II.C.

\noindent \textbf{Wireless Nodes:}
Each wireless node is a Linux PC equipped with an Intel WLAN 8265/8275 wireless network adapter supporting 802.11 standards.

\noindent \textbf{OnePlus 8T:} 
Each smartphone is equipped with the Qualcomm FastConnect 6900 Wi-Fi chipset. The initial locations of the smartphones are illustrated in Fig. 2. During our experiments, the smartphones were placed in various locations within the lab at random intervals.

\noindent \textbf{Wireless Access Point:} 
In this study, we utilized the \textit{ASUS RT-AX88U Pro} as the AP. This device supports various wireless standards, including 802.11, and operates simultaneously in dual-band (2.4 GHz and 5 GHz) modes. It features 2x2 antennas for enhanced beamforming and supports channel capacities ranging from 20 MHz to 160 MHz, with a maximum output power of 20 dBm.

\noindent \textbf{USRP X310:} 
For generating and transmitting malicious interference signals, we used the USRP X310
radio from Ettus Research. This open-source SDR platform features two extended-bandwidth daughterboard slots, covering a frequency range from 10 MHz to 6 GHz. It is equipped with two individually configurable RF channels, each operating at a maximum sample rate of 200 Msps and an effective bandwidth of 160 MHz, with a maximum output power exceeding 20 dBm. The USRP is connected to a workstation running Ubuntu 20.04 via a 10G Ethernet interface, which hosts GNU Radio. This software is utilized to program the USRP for the various jamming attack types implemented in our experiments. Additional details are provided in Section II.C.

\noindent \textbf{Edge Server:}
For the training and inference of our proposed framework, we utilized a customized edge server equipped with an AMD EPYC 7352 2.3 GHz 24-core processor, 128 GB of DDR4 RAM, and four NVIDIA RTX A5000 GPUs, each with 24 GB of memory.

    
    

\subsection{Software Setup}

Before implementing over-the-air attacks, we first configured the AP to operate in a specific band. We disabled the \textit{smart connect} and \textit{agile multiband} options in the AP user interface to ensure operation in either the 2.4 GHz or 5 GHz band. For this work, we selected the 2.4 GHz band due to its limited number of non-overlapping user channels and smaller capacity, allowing our attacks to be more effective with low power and sample rates. However, our approach is adaptable to any frequency range, from 10 MHz to 6 GHz, as supported by our SDR. Next, we identified the channel used by our AP to receive and transmit 802.11 frames using a wireless network assessment tool called \textit{NetSpot}. The analysis revealed that the AP operates on channel 6, with a center frequency of 2.437 GHz and a channel bandwidth of 20 MHz. We also detected neighboring 802.11 APs operating in the same frequency band; however, their interference effects were negligible. To further minimize unintentional interference, all experiments were conducted late at night to reduce the impact from nearby 802.11 networks.


\noindent \textbf{Traffic Simulator:} 
To generate traffic between the wireless nodes and the AP, and thus create a realistic
environment where the nodes communicate during a malicious jamming attack, we created a script written in Python to serve as a \textit{traffic simulator}, automating the traffic exchange between the AP and the nodes. We utilized an open-source network tool called \textit{iperf3} to create UDP data streams between the two ends
in both directions at a rate of 1 Mbps with a packet size of 1000 bytes. The traffic simulator was deployed in the edge server as a background process.

\noindent \textbf{Cross-Layer Collector:} 
To effectively capture network metrics that can indicate a potential jamming attack in real-time, we created a cross-layer component
to collect diverse metrics from different protocol layers (physical, data link, and application layer). This software-based component was implemented using Python and deployed on the edge server as a background process to forward the collected metrics to our ML pipeline for further pre-processing. More specifically, the cross-layer component includes four parts: First, it initiates connections with the AP and the wireless nodes using the SSH protocol. Second, it gathers physical and link layer metrics from each 802.11 device driver. Note that 802.11 device drivers do not offer application layer metrics. Third, it sends probing UDP packets (to overcome the previous issue) with a size of 64 bytes between the wireless nodes to measure and extract application layer metrics. Finally, it combines all of these metrics into a feature vector and forwards it to our ML pipeline for further preprocessing. The duration of this cycle is fully customizable; in our case, each measurement is sampled every 0.5 seconds.

\subsection{Threat Model and Scenarios}

In this work, we assume that a powerful jammer
disrupts all types of communications within a particular frequency range, affecting all three 802.11 channels (channels 1, 6, and 11) in the 2.4 GHz band by broadcasting Additive White Gaussian Noise (AWGN). Additionally, we assume that this malicious signal disrupts communications among nodes within the jamming radius, partially disabling their ability to communicate. However, it does not affect the Cross-Layer Collector’s function to gather and forward the metrics to the edge server. Based on these assumptions, we implemented three types of jammers in GNU Radio: constant jammer, random jammer, and reactive jammer. For each jammer, we experimented with different parameters. First, we controlled the output power of each jammer by adjusting the gain of the USRP from 10 to 30 dBi. Second, instead of injecting only AWGN into the channel, we experimented with different interference signals such as single-tone cosine and sine waves, triangle waves, pulses, and sawtooth waves.

\noindent \textbf{1) Constant Jammer:} In this type of attack, we continuously injected an interference signal into the wireless channel, overwhelming the three non-overlapping channels in the 2.4 GHz band for an extended period of time.

\noindent \textbf{2) Random Jammer:}  In this type of attack, we injected  interference signals at random intervals, overwhelming the wireless channels.

\noindent \textbf{3) Reactive Jammer:}  In this type of attack, we injected  interference signals, only when the energy on the channels exceeded a predefined threshold. The reactive jammer transmitted upon detecting energy above -65 dBm, a setting designed to ensure high sensitivity and minimize false detections as followed in [9]. This threshold also prevented the jammer from responding to signals from nearby 802.11 networks.

\begin{figure}[htbp]
  \centering
  \begin{minipage}{0.49\linewidth}
    \centering
    \includegraphics[width=\textwidth, height=4cm]{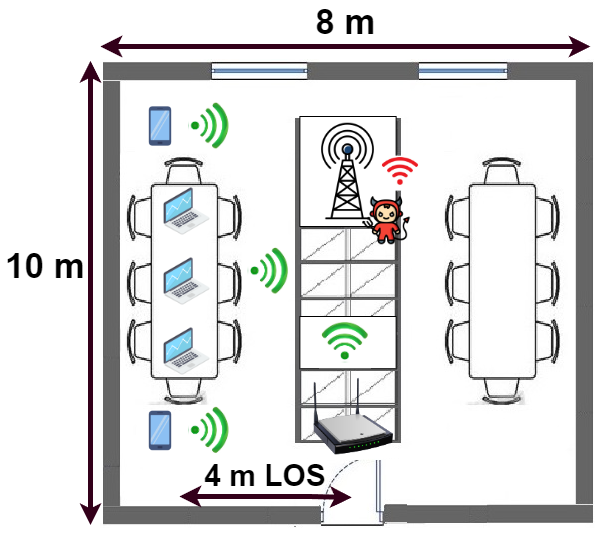} 
    \makebox[3.5cm]{\footnotesize (a) LOS scenario} 
    \label{fig:los}
  \end{minipage}%
  \hspace{-0.1cm} 
  \begin{minipage}{0.49\linewidth}
    \centering
    \includegraphics[width=\textwidth, height=4cm]{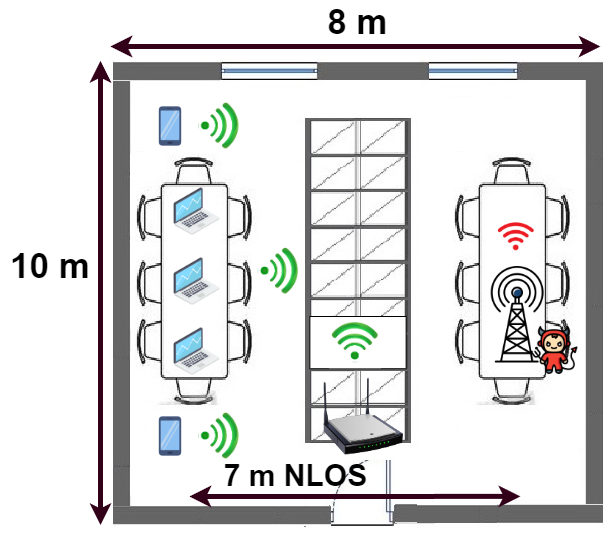} 
    \makebox[3.5cm]{\footnotesize (b) NLOS scenario} 
    \label{fig:nlos}
  \end{minipage}
  \caption{Illustration of the two scenarios.} 
  \label{fig:scenarios} 
\end{figure}

\noindent \textbf{Scenarios:} 
Finally, to examine both ideal and challenging link characteristics, we considered two different configurations, referred to as the Line Of Sight (LOS) scenario and the Non-Line Of Sight (NLOS) scenario, as illustrated in Fig. 2. In the first scenario, shown in (a), we deployed the jammer in the middle of the lab at a height of 3 m, with the wireless nodes positioned around the lab at a distance of 6 m and at a height of 1m. The jammer maintained an LOS with the wireless nodes, allowing it to generate interference signals that propagated directly to them without obstruction or multipath reflection. In the second scenario, the jammer was placed in a different location, as shown in (b), at a height of 1m. To create an NLOS condition, we obstructed the jammer’s signals with a plastic surface and reduced the transmit power by adding attenuation elements at the output of the radio front-end.


\section{JamShield Detection System}
\begin{figure}[!htbp] 
    \centering 
    \includegraphics[width=0.48\textwidth]{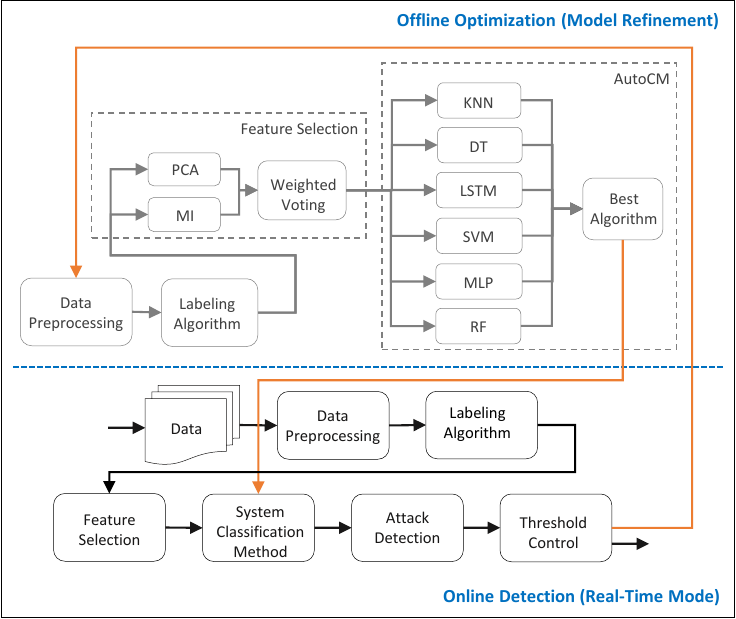} 
    \caption{The proposed detection system architecture.} 
    \label{fig:system} 
\end{figure}

We designed a robust jamming attack detection system called \textit{JamShield} in real-time and offline optimization modes. The system architecture is shown in Fig.~\ref{fig:system}. This architecture integrates different ML techniques to accurately and efficiently identify jamming attacks in communication networks.

%

\noindent The system employs a hybrid feature selection module,  to select the most impactful features. This ensures that only the most relevant and informative features are chosen, optimizing the system’s classification capabilities while reducing dimensionality.
Our architecture is divided into two operational phases:

\noindent \textbf{Online Detection (Real-Time Mode):} During real-time operation, the system continuously processes incoming network features to detect potential jamming attacks. The raw data is preprocessed and passed through the feature selection module. Then, the system classification algorithm analyzes the data to detect potential attacks.  After the detection module, the system checks the threshold parameters to maintain high accuracy through the threshold control module. If one of the performance metrics is lower than its threshold, then the offline optimization mode is activated to decide the best classification algorithm for the current network data. Otherwise, the system operates its current classification method.

\noindent \textbf{Offline Optimization (Model Refinement):} In this mode, the system focuses on improving performance by processing current collected network features. After data preprocessing, the labeling algorithm from \cite{b14} is applied using expectation-maximization and K-means clustering to categorize the data. Then, the feature selection module is activated to identify the most relevant features. These selected features are passed to the AutoCM, which evaluates multiple classification algorithms. After that, the best algorithm module determines the optimal model. The system is then updated with the best-performing algorithm for real-time mode.

\subsection{Feature Selection Module}
This module is designed to identify the most important features from an initial set of 40 features collected from our testbed, while reducing the dataset's dimensionality to ensure faster processing without degrading performance. This module utilises two main techniques: PCA and MI. The results of both methods are combined through a weighted voting mechanism, which ultimately reduces the feature set to the most significant 20 features for jamming attack classification \cite{b16}, \cite{dataport}. This module operates in both online and offline modes to maintain adaptive and optimal feature selection.

\noindent \textit{PCA:} is employed to reduce the dimensionality of the dataset by transforming the original 40 features into a smaller set of key principal components. These components capture the essential patterns within the dataset, removing redundant or noisy features. By retaining only the top principal components, we narrowed the feature set down to 20 features, ensuring that the most critical information is preserved for accurate detection while minimizing unnecessary data.

\noindent \textit{MI:} assesses how each feature relates to the target variable (attack presence), identifying the most significant contributors to the detection task. MI refines the feature set by selecting statistically relevant features for predicting jamming attacks, ensuring accurate and reliable system predictions. 

\noindent The final feature selection is achieved through a \textit{weighted voting} process that combines the results from PCA and MI. Each feature is assigned a weight based on its importance, as determined by PCA, which accounts for variance in the data, and MI, which focuses on prediction accuracy. This weighted voting mechanism prioritizes features that are both varied and predictive, ultimately reducing the feature set from 40 to 20, optimizing it for jamming attack detection.
Our feature selection approach provides several advantages. PCA reduces the number of features, allowing the system to process data more quickly without losing key information. MI ensures the chosen features are highly relevant, making more accurate predictions. By combining these techniques through weighted voting, we achieve a balanced and optimized feature set that enhances the system's precision and efficiency in detecting jamming attacks in real-time.

\subsection{AutoCM Module}
This module is responsible for determining the optimal classification algorithm for real-time detection of jamming attacks. It operates mainly in offline mode to reduce the real-time computational load during real-time operations, where various classification algorithms are tested and evaluated. The module is activated whenever the performance of the current system classification method drops below predefined thresholds. By evaluating performance, the AutoCM ensures that the system adapts dynamically to changing network conditions and maintains stable detection accuracy.

The AutoCM Module includes six classification algorithms: K-Nearest Neighbors (KNN), Decision Tree (DT), Long Short Term Memory (LSTM), Support Vector Machine (SVM), Multi-Layer Perceptron (MLP), and Random Forest (RF). Each algorithm is chosen based on a comprehensive literature review of existing jamming attack detection algorithms. The module dynamically tests these algorithms and selects the one that offers the best balance between speed and accuracy, ensuring optimal real-time performance under dynamic network conditions.

We utilized the threshold control module from our previous work \cite{b15}. The threshold control monitors critical performance metrics such as True Positive (TP) and False Negative (FN) rates to evaluate the sensitivity of the currently active classifier. Specifically, the system computes the sensitivity, $Se$, as:
\begin{equation}
    Se = {\frac{TP}{FN+TP}}
\end{equation}

\begin{table}[htbp]
\centering
\caption{Number of Records for Training/Testing.\label{tab:dataset}}
\begin{tabular}{|l|l|c|}
\hline
\multicolumn{2}{|c|}{\textbf{Class}}  & \textbf{Used Records} \\ \hline\hline
\multicolumn{2}{|l|}{Data Benign} & 29,896 \\ \hline\hline
\multirow{6}{*}{Constant Jammer} 
    & Gaussian 10 dB & 624 \\ \cline{2-3}
    & Gaussian 20 dB & 625 \\ \cline{2-3}
    & Gaussian 25 dB & 624 \\ \cline{2-3}
    & Gaussian Dynamic Gain & 455 \\ \cline{2-3}
    & Pulse 20 dB & 232 \\ \cline{2-3}
    & Triangle 20 dB & 493 \\ \hline\hline
\multirow{6}{*}{Reactive Jammer} 
    & Cos NLOS & 509 \\ \cline{2-3}
    & Gaussian Additional Devices & 537 \\ \cline{2-3}
    & Gaussian LOS & 1,151 \\ \cline{2-3}
    & Gaussian NLOS & 524 \\ \cline{2-3}
    & Square NLOS & 752 \\ \cline{2-3}
    & Triangle NLOS & 531 \\ \hline \hline
\multirow{4}{*}{Random Jammer} 
    & Cos Dynamic Gain & 559 \\ \cline{2-3}
    & Pulse Dynamic Gain & 680 \\ \cline{2-3}
    & Saw Tooth Dynamic Gain & 1,128 \\ \cline{2-3}
    & Triangle Dynamic Gain & 541 \\ \hline
\end{tabular}
\end{table}


If the sensitivity $Se$ falls below a predefined threshold $T$, the system recognizes this as an indication that the current classifier is under-performing.
The threshold ($T$) values for each classifier are as follows: $T_{KNN} = 0.91$, $T_{DT} = 0.925$, $T_{LSTM} = 0.905$, $T_{SVM} = 0.90$, $T_{MLP} = 0.915$, and $T_{RF} = 0.91$.
By applying these predefined threshold values, the system ensures classifiers are dynamically updated whenever their performance falls below these acceptable levels, preventing extended periods of low detection accuracy.

\section{Performance Evaluation}
To train JamShield, we utilize our collected dataset \cite{b16}, \cite{dataport}, which is reduced to 20 features following the dimensionality reduction process outlined in the feature selection section. The dataset distribution features a higher proportion of benign data, constituting roughly three times the number of malicious samples, as detailed in Table~\ref{tab:dataset}. To train and evaluate our proposed detection system, we first divide the dataset into 70\% for training and 30\% for testing, and we scale the features to ensure that all input variables contribute equally to the model's performance. Next, we employ 10-fold cross-validation to train all detection algorithms in the JamShield system for a binary classification task, determining whether a given sample is an attack or benign.

\begin{table}[htbp]
\centering
\caption{Optimized AutoCM Classifier Hyperparameters.\label{tab:hyperparameters}}
\begin{tabular}{|l|l|}
\hline
    \textbf{Classifier} & \textbf{Hyperparameters} \\ \hline\hline

    \textbf{KNN} & 
    \begin{tabular}[c]{@{}l@{}} 
    - Number of Neighbors: 10 \\
    - Weight: Distance \\
    - Metric: Euclidean 
    \end{tabular} \\ \hline\hline

    \textbf{DT} & 
    \begin{tabular}[c]{@{}l@{}} 
    - Max Depth: 15 \\
    - Min Samples Split: 10 \\
    - Criterion: Entropy 
    \end{tabular} \\ \hline\hline

    \textbf{LSTM} & 
    \begin{tabular}[c]{@{}l@{}} 
    - Number of Layers: 2 \\
    - Hidden Units per Layer: 50 \\
    - Learning Rate: 0.001 \\
    - Batch Size: 128 \\
    - Loss: Cross-Entropy
    \end{tabular} \\ \hline\hline
            
    \textbf{SVM} & 
    \begin{tabular}[c]{@{}l@{}} 
    - Kernel: RBF \\
    - Regularization Parameter: 1.0 \\
    - Gamma: Scale 
    \end{tabular} \\ \hline\hline
    
    \textbf{MLP} & 
    \begin{tabular}[c]{@{}l@{}} 
    - Hidden Layers: 3 (100, 50, 25) \\
    - Activation Function: ReLU \\
    - Last Layer: Softmax
    - Learning Rate: 0.01 \\
    - Batch Size: 128 \\
    - Loss: Cross-Entropy
    \end{tabular} \\ \hline\hline

    \textbf{RF} & 
    \begin{tabular}[c]{@{}l@{}} 
    - Number of Trees: 150 \\
    - Max Depth: 20 \\
    - Min Samples Split: 5 \\
    - Criterion: Gini 
    \end{tabular} \\ \hline
\end{tabular}
\end{table}

\begin{figure}[htbp]
    \centering
    \includegraphics[width=3.5in]{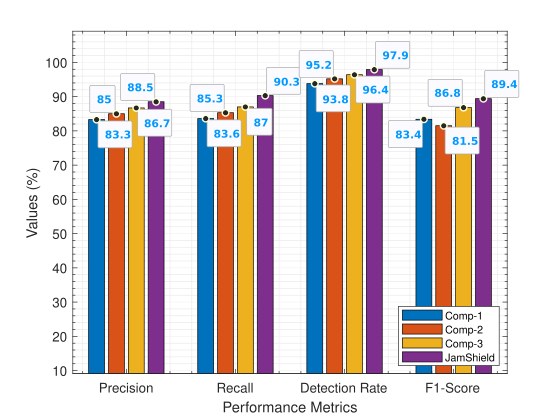}
    \caption{The comparison of performance metrics.}
    \label{fig:performance}
\end{figure}

\begin{figure}[htbp]
    \centering
    \includegraphics[width=3.2in]{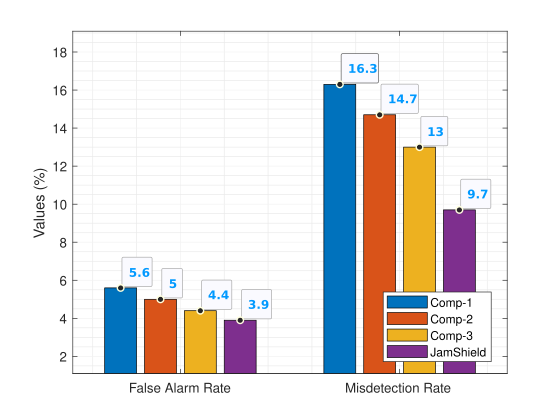}
    \caption{The comparison of error rates.}
    \label{fig:error}
\end{figure}

To further enhance the performance of the JamShield, we conduct hyperparameter tuning on each of the six ML algorithms to optimize their configurations. The optimal parameters and training configurations for each algorithm are presented in Table~\ref{tab:hyperparameters}.
To evaluate the performance of our system, we compare it with three state-of-the-art detection algorithms from the literature.  The first comparison model \cite{b11}, called \textbf{Comp-1}, is a fully connected MLP that utilizes three hidden layers with 453, 207, and 374 neurons, respectively, with ReLU activation functions followed by a softmax layer. The second model \cite{b9}, \textbf{Comp-2}, consists of a fully connected MLP with two hidden layers size of 128 neurons and a stacked kernelized SVM that utilizes an RBF kernel for nonlinear data separation.  The third model \cite{b10}, \textbf{Comp-3}, is a deep learning model with five hidden layers, each containing 1,000 neurons, utilizing the ReLU activation function and followed by a softmax layer. It includes a dropout rate of 0.3 to prevent overfitting, a learning rate of 0.01, and a batch size of 64.

Given the dataset's imbalanced nature, we focus on precision, recall, and F1 score as key metrics for assessing performance. The key performance metrics are shown in Fig.~\ref{fig:performance} for each model. Our solution consistently outperformed the others, achieving the highest Precision (88.5\%), Recall (90.3\%), Detection Rate (97.9\%), and F1-Score (89.4\%), indicating its superior ability to detect and classify jamming attacks accurately. In addition, we evaluated the false alarm rate and the misdetection rate for each model, as shown in Fig.~\ref{fig:error}. 
JamShield outperformed the comparison models in both categories, with the lowest false alarm rate at 3.9\% and misdetection rate at 9.7\%. These results demonstrate that JamShield detects attacks with high accuracy and minimises false positives and misdetections, enhancing its reliability for real-time deployment in communication networks.

JamShield achieves superior performance through its two core modules: (i) a hybrid feature selection mechanism that optimizes classification by reducing dimensionality while retaining discriminative power and (ii) the AutoCM module, which dynamically switches classifiers based on real-time performance thresholds, ensuring adaptability to varying network conditions.
By leveraging PCA and MI, our feature selection process reduces redundant data by approximately 56\% without compromising detection accuracy. This optimization eliminates redundant data, ensuring that computational complexity remains low without sacrificing detection performance. Meanwhile, the AutoCM module enables dynamic classifier selection, adapting in real-time based on network conditions. Unlike static detection models, JamShield continuously monitors its performance and switches to a more effective classifier when necessary, maintaining high detection accuracy even in evolving attack scenarios.
While these enhancements introduce minor computational overhead, they are efficiently managed. The AutoCM module ensures consistent accuracy across diverse conditions with minimal processing cost. JamShield remains suitable for real-time deployment, achieving efficient inference speeds within acceptable limits, balancing adaptability and efficiency for robust jamming detection.

\section{Conclusion}
In this paper, we introduced JamShield, a dynamic jamming detection system designed for real-world wireless environments. It addressed the limitations of existing models through a hybrid feature selection approach and the AutoCM module, ensuring optimal detection performance. Trained on our publicly released over-the-air dataset, JamShield demonstrated significant improvements in detection rate, precision, and recall, with reduced false alarms. These enhancements make JamShield a practical and robust solution for real-world jamming attack detection in wireless networks.

\section*{Acknowledgment}
This work was partially supported by The Scientific and Technological Research Council of Turkey (TUBITAK) 1515 Frontier R\&D Laboratories Support Program for BTS Advanced AI Hub: BTS Autonomous Networks and Data Innovation Lab (Project 5239903), the U.S. Army Research Office (ARO) under grants W911NF-23-1-0088 and DURIP W911NF-23-1-0064, and the National Science Foundation (NSF) RINGS program under grant CNS-2146838.

\input{conference_101719.bbl}

\end{document}